\documentstyle[preprint,aps,epsf,floats]{revtex}

\tighten 

\def\OMIT#1{{}}

\def\Dslash{D\hskip-0.65em /}

\begin{document}

\preprint{\vbox{
\hbox{NT@UW-019}
\hbox{MIT-CTP 3405}
}}

\title{Soft Pion Emission in DVCS}
\author{{\bf Jiunn-Wei Chen}$^{a,b}$,
and {\bf Martin J.~Savage}$^{c}$}
\address{$^a$ 
Department of Physics, National Taiwan University, 
\\
Taipei 10617, Taiwan.\\}
\address{$^b$ Center for Theoretical Physics, Massachusetts Institute of 
Technology, \\
Cambridge, MA 02139.\\
}
\address{$^c$ Department of Physics, University of Washington, \\
Seattle, WA 98195.\\ }

\maketitle

\begin{abstract} 
We examine soft-pion emission in  deeply virtual Compton
scattering.
Contrary to previous claims, we find that the amplitude 
for soft-pion emission
is not directly related to the generalized parton 
distributions in the nucleon.
\end{abstract}

\bigskip
\vskip 8.0cm
\leftline{August 2003}

\vfill\eject

\section{Introduction}

Deep inelastic scattering (DIS) from the nucleon, 
$e N\rightarrow e^\prime X$, has played a central role in
the development of our understanding of QCD.
Application of the optical theorem to the inclusive cross-section for DIS
relates it to the forward matrix element of the time-ordered product of two
electromagnetic currents.
As QCD is  asymptotically free at short-distances, an operator product
expansion of the time-ordered product can be performed which allows the total
inclusive cross section to be related to the forward matrix elements 
in the nucleon of twist-2 operators involving the quark and gluon fields.
One particularly nice implication of this is that 
these forward matrix elements correspond to the moments of 
the parton distribution functions (PDF's) in the nucleon,
where the PDF's are probability distributions.
Recently, it has been realized that the off-forward matrix elements of these
twist-2 operators are also of interest~\cite{Ji:1996ek,Ji:1996nm,Dittes:xz,Muller:1998fv,Collins:1996fb,Radyushkin:1996ru,Radyushkin:1997ki,Radyushkin:1998es}.
Of particular note is the relation between the off-forward matrix elements of
the quark and gluon energy-momentum tensors and the fraction of the 
total angular momentum of the nucleon 
carried by the quarks and gluons~\cite{Ji:1996ek,Ji:1996nm}.
Away from the forward direction the idea of a 
generalized parton distribution (GPD) has been introduced to provide a
somewhat unified description of the scattering 
processes~\cite{Ji:1996ek,Ji:1996nm,Dittes:xz,Muller:1998fv,Collins:1996fb,Radyushkin:1996ru,Radyushkin:1997ki,Radyushkin:1998es}.
While the forward limit of the GPD's reproduce the PDF's, 
weighted integration of the GPD's over Bjorken $x$ 
reproduce the form-factors of the twist-2 operators in the nucleon
(for a nice review see Filippone and Ji~\cite{Filippone:2001ux}).
Gravitational scattering is not a feasible option for the measurement of the 
matrix element of the energy-momentum tensor,
however, deeply virtual Compton scattering
(DVCS)~\cite{Ji:1996nm,Ji:1998xh,Radyushkin:1996nd,Collins:1998be} 
provides an experimentally realized 
alternate process with which to explore the matrix elements
of the twist-2 operators over a wide range of kinematics.

A significant experimental effort has been mounted to measure the GPD's through
DVCS~\cite{Airapetian:2001yk,Stepanyan:2001sm,d'Hose:qw,Favart:2002eh}. 
In general, each GPD has three kinematic arguments, $x$, $\xi$ and $t$,
e.g. $H^q(x,\xi,t)$.
Elastic DVCS with high energy electrons is determined by a convolution 
in the variable $x$ of the GPD's with $1/(\pm x-\xi + i\epsilon)$,
and therefore one cannot determine the GPD's directly from such measurements.
However, at lower energies and with polarized beams one can 
determine the imaginary part of this convolution which gives 
the GPD's evaluated at $x=\xi$,
e.g. Ref.~\cite{d'Hose:qw}.
Soft-pion emission during DVCS 
$\gamma^* p\rightarrow\gamma p \pi$ may provide 
a background to the elastic process $\gamma^* p\rightarrow\gamma p$ 
due to limited detector resolution in some experiments.
The matrix elements and cross section for soft-pion emission during 
DVCS (DVCS-pion production) have been computed by 
Guichon, Mosse and Vanderhaeghen~\cite{Guichon:2003ah} using current algebra.
They find that there are no further unknown strong interaction contributions
beyond the GPD's of the nucleon.
In this paper we compute the amplitude for soft-pion emission during DVCS
(DVCS-pion production)
using chiral perturbation theory.
We find that at leading order in the chiral expansion, the amplitude for  
DVCS-pion production is controlled by the PDF's and by additional
strong interaction 
parameters that are not related to the elastic GPD's.  
This result disagrees with the result of Ref.~\cite{Guichon:2003ah}.

\section{Twist-2 Matrix Elements and Chiral Perturbation Theory}

The leading order chiral Lagrangian describing the low-energy dynamics of the 
nucleons, $\Delta$'s and pions (pseudo-Goldstone bosons)
that is consistent with the spontaneously broken
$SU(2)_L\otimes SU(2)_R$ approximate chiral symmetry of QCD 
is~\cite{Jenkins:1990jv,Jenkins:1991ne}
\begin{eqnarray}
{\cal L} & = & 
\overline{N}\ iv\cdot {\cal D} \ N\ - \ 
\overline{T}_\mu\ iv\cdot {\cal D}\  T^\mu
\ +\ 
\Delta\ \overline{T}_\mu\ T^\mu
\nonumber\\
&& \ +\ 
{f^2\over 8} {\rm Tr}\left[ \partial_\mu\Sigma^\dagger \ \partial^\mu\Sigma\
\right]
\ +\ \lambda\ {f^2\over 4}  {\rm Tr}\left[ m_q \Sigma^\dagger\ +\ {\rm h.c.}\
\right]
\nonumber\\
&& \ +\ 
2 g_A\  \overline{N} S^\mu  A_\mu N
\ +\  
g_{\Delta N}\ 
\left[\ 
\overline{T}^{abc,\nu}\  A^d_{a,\nu}\  N_b \ \epsilon_{cd} 
\ +\ {\rm h.c.}
\ \right]
\ +\ 
2 g_{\Delta\Delta}\  \overline{T}_\nu S^\mu  A_\mu T^\nu
\ \ \ ,
\label{eq:lagstrong}
\end{eqnarray}
where ${\cal D}$ is the chiral covariant derivative, $N$ is the nucleon field
operator, and $T^\mu$ is the Rarita-Schwinger
field containing the quartet of spin-${3\over 2}$ $\Delta$-resonances.
The mass difference between the $\Delta$-resonances and the nucleon
is $\Delta$, and 
the axial couplings between the baryons and pions are
$g_A$, $g_{\Delta N}$ and $g_{\Delta\Delta}$.
$S_\mu$ is the covariant spin vector defined in heavy-baryon 
$\chi$PT~\cite{Jenkins:1990jv,Jenkins:1991ne}, and 
$v_\mu$ is the heavy-baryon four-vector, with $v^2=1$.
The pions appear through $\Sigma$ and $A_\mu$ which are defined to be
\begin{eqnarray}
\Sigma & = & \exp\left({2 i M \over f}\right)
\ =\ \xi^2
\ ,\ 
A^\mu \ =\  {i\over 2}\left(\ \xi\partial^\mu\xi^\dagger
-
\xi^\dagger\partial^\mu\xi \ \right)
\ ,\ 
M \ = \ \left(\matrix{\pi^0/\sqrt{2} & \pi^+\cr \pi^- & -\pi^0/\sqrt{2}}\right)
\ .
\label{eq:mesons}
\end{eqnarray}
We have not shown flavor indices in eq.~(\ref{eq:lagstrong})
for terms where the contractions are unambiguous.
Under chiral transformations the various fields transform as
\begin{eqnarray}
&&
\Sigma\rightarrow L\Sigma R^\dagger
\ \ ,\ \ 
N\rightarrow UN
\ \ ,\ \ 
T^\mu\rightarrow UUU T^\mu
\ \ ,\ \ 
A^\mu\rightarrow UA^\mu U^\dagger
\ \ \ ,
\label{eq:trans}
\end{eqnarray}
where $L$ and $R$ denote left- and right-handed chiral transformations
respectively.
The $\xi$ field 
(not to be confused with the kinematic variable $\xi$ 
that appears as an argument of the GPD's)
transforms as $L\xi U^\dagger = U\xi R^\dagger$, which defines
the transformation $U$.
The mass-matrix, which is treated as a spurion field, 
is taken to transform as $m_q\rightarrow L m_q R^\dagger$.

The twist-2 operators involving the quark fields
that provide the dominant contribution to DIS and DVCS are
\begin{eqnarray}
\theta_{V,\ \mu_1 ..\mu_n }^{(n),0} & = &  (i)^{n-1}\ 
\overline{q} \ \gamma_{\{ \mu_1}\ 
\stackrel{\leftrightarrow}{D}_{\mu_2}\ 
...\ 
\stackrel{\leftrightarrow}{D}_{\mu_n\} }\ q
\nonumber\\
\theta_{V,\ \mu_1 ..\mu_n }^{(n),b} & = & (i)^{n-1}\ 
\overline{q} \ \gamma_{\{ \mu_1}\ 
\stackrel{\leftrightarrow}{D}_{\mu_2}\ 
...\ 
\stackrel{\leftrightarrow}{D}_{\mu_n\} }\ \tau^b\ q
\nonumber\\
\theta_{A,\ \mu_1 ..\mu_n }^{(n),0} & = & (i)^{n-1}\ 
\overline{q} \ \gamma_{\{ \mu_1}\ 
\stackrel{\leftrightarrow}{D}_{\mu_2}\ 
...\ 
\stackrel{\leftrightarrow}{D}_{\mu_n\} }\ \gamma_5\ q
\nonumber\\
\theta_{A,\ \mu_1 ..\mu_n }^{(n),b} & = & (i)^{n-1}\ 
\overline{q} \ \gamma_{\{ \mu_1}\ 
\stackrel{\leftrightarrow}{D}_{\mu_2}\ 
...\ 
\stackrel{\leftrightarrow}{D}_{\mu_n\} }\ \gamma_5\ \tau^b\ q
\ \ \ ,
\label{eq:twisttwo}
\end{eqnarray}
where the indices enclosed by $\{ ... \}$ are symmetrized and their traces
are removed.
The operator $\stackrel{\leftrightarrow}{D}_{\mu }$ is 
$\stackrel{\leftrightarrow}{D}_{\mu } =
\overrightarrow{D}_\mu-\overleftarrow{D}_\mu$,
where $D_\mu=\partial_\mu + i g A_\mu$ is the QCD covariant derivative, and $g$
is the strong coupling constant.
Hadronic matrix elements of the operators in eq.~(\ref{eq:twisttwo}) 
in the kinematic regimes appropriate for DIS and slightly off-forward DVCS
can be described in chiral perturbation theory 
($\chi$PT)~\cite{Arndt:2001ye,Chen:2001eg,Chen:2001et,Chen:2001nb,Chen:2001pv,Belitsky:2002jp}
as the momentum transfer to the hadronic system is small compared 
to the scale of chiral symmetry breaking, $\Lambda_\chi\sim 1~{\rm GeV}$.

Matrix elements of the isoscalar operators 
$\theta_{V,\ \mu_1 ..\mu_n }^{(n),0}$ and 
$\theta_{A,\ \mu_1 ..\mu_n }^{(n),0}$
in the nucleon
are reproduced at leading order in the chiral expansion by the 
operators~\cite{Arndt:2001ye,Chen:2001eg}
\begin{eqnarray}
\theta_{V,\ \mu_1 ..\mu_n }^{(n),0} & \rightarrow & 
M^{n-1}\ 
\langle x^{n-1}\rangle_{ q_V^{(0)}}\ \ 
v_{\mu_1} v_{\mu_2} \ ...\ v_{\mu_n} \ \ 
\overline{N} N
\nonumber\\
\theta_{A,\ \mu_1 ..\mu_n }^{(n),0} & \rightarrow & 
M^{n-1}\ 
\langle x^{n-1}\rangle_{ q_A^{(0)}}\ \ 
\ v_{ \{ \mu_1} v_{\mu_2} \ ...\ v_{\mu_{n-1}} \ \ 
\overline{N} \ S_{\mu_n \} }\ N
\ \ \ ,
\label{eq:isoscalar}
\end{eqnarray}
where $\langle x^{p}\rangle_{ q_V^{(0)}}$ and 
$\langle x^{p}\rangle_{ q_A^{(0)}}$
are the $p$'th moments of the 
isoscalar vector and axial-vector PDF's in the nucleon, respectively.
$M$ is the nucleon mass.

To construct matrix elements of the isovector operators one introduces the
spurion fields
$\tau^a_L$ and $\tau^a_R$ which transform as
\begin{eqnarray}
\tau^a_L & \rightarrow & L \tau^a_L L^\dagger
\ \ ,\ \ 
\tau^a_R \ \rightarrow \ R \tau^a_R R^\dagger
\ \ \ ,
\end{eqnarray}
under $SU(2)_L\otimes SU(2)_R$
and whose vevs are $\langle \tau^a_L \rangle = \langle \tau^a_R \rangle = 
\tau^a$.  It is then convenient to define
\begin{eqnarray}
\tau^b_{\xi,\pm}
& = & {1\over 2}
\left(\ \xi^\dagger\tau^b_L\xi \pm \xi\tau^b_R\xi^\dagger\ \right)
\ \ \ ,
\end{eqnarray}
that transforms as $\tau^b_{\xi,\pm}\rightarrow U \tau^b_{\xi,\pm} U^\dagger$
under $SU(2)_L\otimes SU(2)_R$, and as 
$\tau^b_{\xi,\pm}\rightarrow \pm \tau^b_{\xi,\pm}$ under parity.
At leading order in the chiral expansion
the matrix elements of the isovector operators
$\theta_{V,\ \mu_1 ..\mu_n }^{(n),b}$ and 
$\theta_{A,\ \mu_1 ..\mu_n }^{(n),b}$ in the nucleon and between the nucleon
and $\Delta$ (required for our calculation of soft-pion emission)
are reproduced by 
\begin{eqnarray}
\theta_{V,\ \mu_1 ..\mu_n }^{(n),b} & \rightarrow & 
M^{n-1}\ 
\langle x^{n-1}\rangle_{ q_V^{(1)}}\ \ 
v_{\mu_1} v_{\mu_2} \ ...\ v_{\mu_n} \ \ 
\overline{N} \ \tau^b_{\xi,+}\ N
\nonumber\\
& &\ +\ 
M^{n-1}\ 
a^{(n)}\ v_{ \{ \mu_1} v_{\mu_2} \ ...\ v_{\mu_{n-1}} \ \ 
\overline{N}\  S_{\mu_n \} }\ \tau^b_{\xi,-}\ N
\nonumber\\
\theta_{A,\ \mu_1 ..\mu_n }^{(n),b} & \rightarrow & 
M^{n-1}\ 
\langle x^{n-1}\rangle_{ q_A^{(1)}}\ \ 
v_{ \{ \mu_1} v_{\mu_2} \ ...\ v_{\mu_{n-1}} \ \ 
\overline{N}\  S_{\mu_n \} }\  \tau^b_{\xi,+}\ N
\nonumber\\
& &\ +\ 
M^{n-1}\ 
b^{(n)}\ 
v_{\mu_1} v_{\mu_2} \ ...\ v_{\mu_n} \ \ 
\overline{N}\ \tau^b_{\xi,-}\ N
\nonumber\\
& &\ +\ 
M^{n-1}\ 
c^{(n)}\ 
\ v_{ \{ \mu_1} v_{\mu_2} \ ...\ v_{\mu_{n-1}} \ \ 
\left[\ 
\overline{T}_{\mu_n \}}^{ijk}\  
 \left(\tau^b_{\xi,+}\right)^l_i \ N_j\ \epsilon_{kl}
\ +\ {\rm h.c.}\ \right]
\ \ \ ,
\label{eq:isovector}
\end{eqnarray}
where 
$\langle x^{p}\rangle_{ q_V^{(1)}}$ and 
$\langle x^{p}\rangle_{ q_A^{(1)}}$ are the $p$'th moments of the isovector
vector and axial-vector PDF's in the nucleon, respectively.
The coefficients $a^{(n)}$, $b^{(n)}$, and $c^{(n)}$ 
cannot be related 
to the PDF's in the nucleon
(and hence the GPD's) by chiral symmetry.
This disagrees with the conclusion of 
Guichon, Mosse and Vanderhaeghen~\cite{Guichon:2003ah} who claim that
chiral symmetry alone is sufficient to relate
$a^{(n)}$ and  $b^{(n)}$ to $\langle x^{n-1}\rangle_{ q_V^{(1)}}$ and 
$\langle x^{n-1}\rangle_{ q_A^{(1)}}$
in the kinematic regime $m_\pi^2\ll -t\ll\Lambda_\chi^2$.

At higher orders in the chiral expansion there will be contributions 
from local operators  that
depend upon $v\cdot q/\Lambda_\chi$ 
(the energy transfer to the hadrons in their rest frame)
and $q^2/\Lambda_\chi^2$ (where $q^2=t$).  
These operators, along with pion loop contributions, 
will contribute to the off forward-matrix elements
and will provide the low-energy behavior of the GPD's.
Further, there will be additional operators whose matrix elements between
single nucleon states vanish, but can contribute to DVCS-pion production.

We have not shown the matrix elements of the twist-2 operators in the pion
sector~\cite{Arndt:2001ye}. For an operator with $n$ indices, the matrix
element between single pion states is suppressed by factors of $p_\pi^n$, where
$p_\pi$ is the pion four-momentum.
For $n\ge 2$ the contribution to DVCS-pion production
from insertions of the twist-2 operators into the pion is a higher order
contribution.

\section{DVCS-pion Production off the Nucleon}

It is straightforward to compute the leading order contribution to 
$\gamma^*N\rightarrow \gamma N\pi$ in $\chi$PT.
Leading order corresponds to ${\cal P}^0$ in the power-counting, 
where ${\cal P}\sim k,q,\Delta,m_\pi$ are the
small expansion parameters, where $k$ is the pion four-momentum and  $q$ is the 
four-momentum injected by the twist-2 operator.
We compute just one of the possible processes, 
$\gamma^* p \rightarrow \gamma n\pi^+$,
in this work and 
the matrix elements for other possible initial and final states can
be constructed analogously.
\begin{figure}[!ht]
\centerline{{\epsfxsize=6.0in \epsfbox{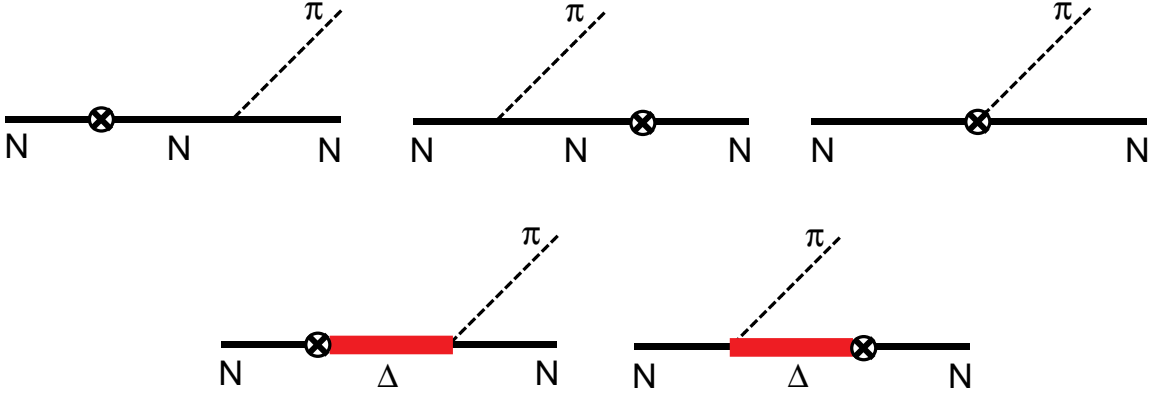}}} 
\vskip 0.15in
\noindent
\caption{\it 
The tree-level diagrams that give the leading order contributions
to DVCS-pion production for $n\ge 2$.
The crossed-circle denotes an insertion of a twist-2 operator.
}
\label{fig:treen2}
\vskip .2in
\end{figure}
At leading order in the chiral expansion, 
the matrix elements for $n\ge 2$  result from the 
diagrams shown in Fig.~\ref{fig:treen2}, and are found to be 
\begin{eqnarray}
\langle n\pi^+| \theta_{V,\ \mu_1 ..\mu_n }^{(n),0} | p\rangle
& = & 0
\nonumber\\
\langle n\pi^+| \theta_{V,\ \mu_1 ..\mu_n }^{(n),3} | p\rangle
& = & 
- i {4 g_A\over f}\ 
M^{n-1}\ 
\langle x^{n-1}\rangle_{ q_V^{(1)}}\ \
{v_{\mu_1} v_{\mu_2} \ ...\ v_{\mu_n} 
\over v\cdot q}\ 
\overline{U}_n \ S\cdot k\ U_p
\nonumber\\
&  & -i M^{n-1}\ 
{2 a^{(n)}\over f}\ 
v_{\{ \mu_1} v_{\mu_2} \ ...\ v_{\mu_{n-1}} \ 
\overline{U}_n \ S_{\mu_n\}}\ U_p
\nonumber\\
\langle n\pi^+| \theta_{A,\ \mu_1 ..\mu_n }^{(n),0} | p\rangle
& = & 
i {2 g_A\over f}\ 
M^{n-1}\ 
\langle x^{n-1}\rangle_{ q_A^{(0)}}\ \
{v_{\{ \mu_1} v_{\mu_2} \ ...\ v_{\mu_{n-1}} 
\over v\cdot q}\ \
i \ \epsilon^{\mu_{n\}}\alpha\lambda\sigma} \ k_\alpha\  v_\lambda\  
\overline{U}_n \ S_\sigma \ U_p
\nonumber\\
\langle n\pi^+| \theta_{A,\ \mu_1 ..\mu_n }^{(n),3} | p\rangle
& = & 
- i {g_A\over f}\ 
M^{n-1}\ 
\langle x^{n-1}\rangle_{ q_A^{(1)}}\ \
{ v_{\{ \mu_1} v_{\mu_2} \ ...\ v_{\mu_{n-1}} \ \
\over v\cdot q}\  \
\left( v\cdot q\ v_{\mu_n \}} - k_{\mu_n \}}\right)
\overline{U}_n \ U_p
\nonumber\\
&  & -
i M^{n-1}\ 
{2 b^{(n)}\over f}\ 
v_{\mu_1} v_{\mu_2} \ ...\ v_{\mu_n} \ 
\overline{U}_n \ U_p
\nonumber\\
&  & +
i M^{n-1}\ 
{8\over 9} {g_{\Delta N} c^{(n)}\over f}
{ v_{\{ \mu_1} v_{\mu_2} \ ...\ v_{\mu_{n-1}} \
 \over  (v\cdot q)^2 - \Delta^2}\ 
\overline{U}_n 
\ \left[\ 
 v\cdot q\ \left( v\cdot q\  v_{\mu_{n}\}} - k_{\mu_{n}\}}\right)
\right.\nonumber\\ &&\left.
\qquad\qquad\qquad\qquad\qquad\qquad\qquad\qquad
\ +\ 
i \ \Delta\  \varepsilon_{\mu_n\} \alpha\sigma\lambda} 
\ k^\alpha \ v^\sigma \ S^\lambda
\right]\  U_p
\ ,
\label{eq:piamps}
\end{eqnarray}
where we have used $v\cdot q=v\cdot k$ in the heavy baryon limit,
and where it is clear that all contributions to these matrix elements are 
${\cal O}({\cal P}^0)$ in the power-counting.
$U_p$ and $U_n$ are the spinors associated with the proton and neutron respectively.
The matrix elements will be modified near the peak of the $\Delta$-pole, where
the $\Delta$-width will need to be resummed into the propagator.
It is important to note that the matrix elements in eq.~(\ref{eq:piamps}) 
are the leading order contribution for all kinematics such that
$(v\cdot q)^2 , |t| , m_\pi^2 \ll \Lambda_\chi^2$ for any value of the
ratio $t/m_\pi^2$.
We do not go on to compute the differential and total cross sections, as this
can be done straightforwardly and does not add anything to our discussion.
The point is that there are additional contributions to the 
DVCS-pion production cross section that are not related to the nucleon 
PDF's (GPD's), 
as encapsulated by the constants $a^{(n)}$ and  $b^{(n)}$.
A similar feature is probably true for any generic DVCS inelastic process, 
where the PDF's or GPD's of the nucleon
can only describe part of the contribution.

\section{Current Conservation and Relations Between Coefficients}

Only the chiral transformation properties of the twist-2 operators
were used  in the construction of the most 
general set of leading order operators in eq.~(\ref{eq:isoscalar}) and
eq.~(\ref{eq:isovector}).
As such, we did not include the fact that the $n=1$ vector operators are
conserved and the $n=1$ axial operators are partially conserved.
These conservation laws lead to additional constraints on the coefficients
$a^{(1)}$ and $b^{(1)}$.
\begin{figure}[!ht]
\vskip -0.2in
\centerline{{\epsfxsize=2.0in \epsfbox{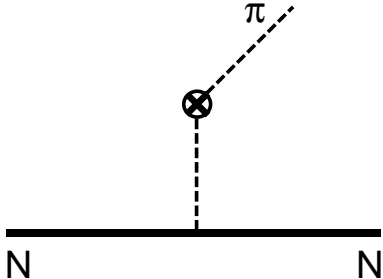}}} 
\vskip 0.15in
\noindent
\caption{\it 
The additional tree-level diagram that contributes at leading order
to the $n=1$ matrix element for DVCS-pion production
of the isovector-vector twist-2 operator
(the isovector-vector current operator).
The crossed-circle denotes an insertion of the $n=1$ twist-2 operator.
}
\label{fig:treen1}
\vskip .2in
\end{figure}
We have only presented the matrix elements for 
$\gamma^* p \rightarrow \gamma n\pi^+$ for $n\ge 2$ in eq.~(\ref{eq:piamps})
as the $n=1$ matrix elements are trivial.
However, it is useful to explore the
impact of current conservation on the $n=1$ vector-isovector matrix element,
which receives an additional contribution at leading order from the 
diagram in Fig.~\ref{fig:treen1}, and is found to be 
\begin{eqnarray}
\langle n\pi^+|\  \theta_{V,\ \mu}^{(1),3}\  | p\rangle
& = & 
- i {4 g_A\over f}\ 
{v_{\mu} \over v\cdot q}\ 
\overline{U}_n \ S\cdot k\ U_p
\ - \ 
i {2 a^{(1)}\over f}\   \overline{U}_n \ S_{\mu}\ U_p
\nonumber\\
& &  - i {4 g_A\over f}\ 
\overline{U}_n { S\cdot (k-q)\ (2k-q)^\mu\over (k-q)^2-m_\pi^2} U_p
\ .
\end{eqnarray}
Conservation of the vector current $\overline{q}\gamma_\mu \tau^a q$ requires
that
$q^\mu \langle n\pi^+|\  \theta_{V,\ \mu}^{(1),3}\  | p\rangle
=0$, and hence
\begin{eqnarray}
q^\mu \langle n\pi^+|\  \theta_{V,\ \mu}^{(1),3}\  | p\rangle
& = & 
-i {2\over f} \left[\ a^{(1)} + 2 g_A\ \right]\ \overline{U}_n \ S\cdot q\ \
U_p
\ =\ 0
\ ,
\end{eqnarray}
from which we conclude that $a^{(1)} = -2 g_A$.
This agrees with the known result derived directly
from the vector current operator in heavy-baryon 
$\chi$PT~\cite{Jenkins:1990jv,Jenkins:1991ne}
and is the relation that was found in Ref.~\cite{Guichon:2003ah}.
In addition, Ref.~\cite{Guichon:2003ah} makes the more general statement that
$a^{(n)} = -2 \langle x^{n-1}\rangle_{ q_A^{(1)}}$,
which reproduces the $n=1$ result as
$\langle 1 \rangle_{ q_A^{(1)}} = g_A$.

Lets consider the implication of having 
$a^{(n)} = -2 \langle x^{n-1}\rangle_{ q_A^{(1)}}$, as claimed
in Ref.~\cite{Guichon:2003ah}.
Specifically, consider the $n=2$ isovector operator
$\theta_{V,\ \mu\nu }^{(2),b} $, which is not conserved
due to interactions with the gluon field, and it is straightforward to
show that in the chiral limit
\begin{eqnarray}
\partial_\mu\ \theta_{V,\ \mu\nu }^{(2),3}
& = & 
-4 \ g\  \overline{q}\ \gamma^\alpha \ G_{\alpha\nu} \ \tau^3\  q
\ \ \ ,
\label{eq:divtwo}
\end{eqnarray}
where $G_{\alpha\beta}$ is the gluon field strength tensor, defined by
$\left[ D_\alpha , D_\beta \right] = i g G_{\alpha\beta}$,
and we have used the equation of motion $\Dslash \ q=0$.
In the effective theory, one finds that
\begin{eqnarray}
-i q^\mu \langle n\pi^+|\  \theta_{V,\ \mu\nu}^{(2),3}\  | p\rangle
& = & 
-{4 g_A\over f} \ M\ 
\langle x\rangle_{ q_V^{(1)}}\ \
v_\nu\ \overline{U}_n\ S\cdot k\ U_p
\nonumber\\
&&
\ -\ M\ 
{a^{(2)}\over f} \ 
\overline{U}_n\ \left( v\cdot q S_\nu + v_\nu S\cdot q \right) U_p
\ \ \ .
\end{eqnarray}
Given that one cannot presently compute the matrix element of 
$\partial_\mu\ \theta_{V,\ \mu\nu }^{(2),3}$, 
and chiral symmetry alone does not relate $a^{(n)}$ to single nucleon matrix
elements, we conclude that
the assertion of Ref.~\cite{Guichon:2003ah} is false.

There are a few assertions in Ref.~\cite{Guichon:2003ah} that have contributed
in some part to their error.
First, the non-local operator that arises at the scale of the scattering, $Q^2$,
was used in all operator manipulations.  The difficulties in dealing with
non-local operators are well-known, and efforts to systematize non-local
effective field theory have shown it to be significantly more 
complex~\cite{Bhansali:1992we} than local effective field theories.
A straightforward problem with this can be seen in  the fact that 
$\theta_{V,\ \nu }^{(1),3}$ and $\theta_{V,\ \mu_1..\mu_{10} }^{(10),3}$
originate from the same non-local operator, but their scale dependence due to
strong-interactions are quite different.
Second, after re-deriving pole-ology (something that has been well understood
for decades)
a central feature of their soft-pion analysis
requires them to work at non-zero quark mass, and they are unable to derive
their result working exclusively in the chiral limit.
Hand waving arguments, including the unjustifiable neglect of 
$\left[ {d\over dt}Q_5 ,  \theta^{(n)}_{V,A} \right]$,
are required in order for them to consider the chiral
limit. 
We believe that these two aspects of the calculation of 
Ref.~\cite{Guichon:2003ah} are flawed.
During the thirty or more years since the techniques used in Ref.~\cite{Guichon:2003ah}
were introduced, significant progress has been made in understanding how to deal
with the approximate chiral symmetry of QCD.  We have used these well-known modern tools
in our analysis.

\section{Conclusions}

We have explored soft-pion production associated with  DVCS.
Using the transformation properties of the 
twist-2 operators under the $SU(2)\otimes SU(2)$ chiral symmetry
we have shown that there are contributions to 
DVCS-pion production that are not related to the GPD's 
of the nucleon, even at
leading order in the chiral expansion.  This contradicts recent claim 
in Ref.~\cite{Guichon:2003ah}.
Further, the contribution from nucleon structure depends upon the PDF's and
one does not need to consider the GPD's at leading order.
Therefore, we conclude that DVCS-pion production does not
depend only on the GPD's of the nucleon, 
and gives rise to a presently incalculable
background to extracting GPD's from DVCS in some experiments.
However, naive dimensional analysis can be used to make a rough estimate 
of the unknown coefficients, and hence a rough estimate of the size of the
background.
It may be the case that a combined study of one-pion and 
two-pion emission during DVCS will 
allow for a determination of the constants $a^{(n)}$, $b^{(n)}$ and 
$c^{(n)}$.  However, there are additional contributions to this process coming
from operators that contribute to $\gamma^* N\rightarrow \gamma \Delta\pi$
but not to $\gamma^* N\rightarrow \gamma \Delta$.  
A significantly 
more comprehensive investigation is required in order to 
ascertain if it is possible to
disentangle all contributions at leading order in the chiral expansion.

\bigskip\bigskip

\acknowledgements

We would like to thank Xiang-Dong Ji, 
David Kaplan and Jerry Miller for useful discussions.
JWC thanks the Institute of Nuclear
Theory at the University of Washington for its hospitality.
JWC is supported, in part, by the National Science Council of ROC and
the U.S. department of energy under grant DOE/ER/40762-213. 
MJS is supported in part by the
U.S. Dept. of Energy under Grant No.~DE-FG03-97ER4014.

\end{document}